\documentclass[12pt,preprint]{aastex}

\usepackage{wasysym}
\newcommand{\rev}{ }

\shorttitle{}
\shortauthors{Dimitri Veras, Alexander J. Mustill, Amy Bonsor}

\begin{document}
\baselineskip 14.pt

\title{The evolution and delivery of rocky extra-solar materials to white dwarfs}

\author{Dimitri Veras$^{1,2,3}$}
\author{Alexander J. Mustill$^{4}$}
\author{Amy Bonsor$^{5}$}
\affil{(1) Centre for Exoplanets and Habitability, University of Warwick, \\ 
Coventry CV4 7AL, United Kingdom}
\affil{(2) Centre for Space Domain Awareness, University of Warwick, \\ 
Coventry CV4 7AL, United Kingdom}
\affil{(3) Department of Physics, University of Warwick, \\ 
Coventry CV4 7AL, United Kingdom}
\affil{(4) Lund Observatory, Department of Astronomy and Theoretical Physics, \\
Lund University, Box 43, 221 00, Lund, Sweden}
\affil{(5) Institute of Astronomy, University of Cambridge, Madingley Road, \\
Cambridge CB3 0HA, UK}

%\begin{abstract} 
%White dwarf 
%\end{abstract}
\keywords{Asteroids, Planets, White Dwarfs, Disks, Geochemistry, Planetary Interiors, Planet Formation, Celestial Mechanics, Tides, Evolved Stars}

\section{Introduction}

The most direct and substantial way to measure the geochemistry of extrasolar objects is to perform chemical autopsies on their broken-up remains. However, traditional ways of probing extrasolar planetary systems around Sun-like stars cannot achieve this goal, instead yielding just a planet's bulk density or chemical details of its atmosphere. Fortunately, new techniques have emerged which allow us to perform the desired geochemical postmortems. These techniques work only on {\it old} planetary systems: specifically, systems which are old enough so that their parent stars have already exhausted their fuel and have evolved into burnt cores known as {\it white dwarfs}.

Hence, understanding stellar evolution (how the star transforms itself over time) and its effect on planetary systems is crucial for correctly interpreting the chemical constraints of exo-planetary material that can be given to us by white dwarfs. The previous two chapters have focussed on the formation of planetary systems and their evolution when they are young and in middle-age. The content in those chapters provide the initial conditions for the late-stage evolution of those systems. The subsequent chapter will then outline the compositional measurements obtained in white dwarf planetary systems.

This chapter will detail the transition from young to old, and describe how asteroids, moons, and comets, as well as boulders, pebbles and dust, evolve into eventual targets for chemical spectroscopy, and how planets and companion stars play a vital role in reshaping system architectures for this purpose. Related reviews from the astrophysical literature which cover these themes include \cite{veras2016a}, \cite{bonxu2017}, \cite{jaccar2018}, \cite{vanlierap2018}, \cite{vanrap2018} and \cite{veras2021}; existing reviews which are focussed more on the content of the subsequent chapter are \cite{juryou2014}, \cite{farihi2016}, \cite{zucyou2018} and \cite{xubon2021}.

For pedagogical purposes, we divide the future evolution of planetary systems from their middle-aged states into three stages, each of which is covered in a separate section in this chapter. The first section (``Stage 1") addresses the changes to a planetary system when a Sun-like star undergoes significant variations in mass, size and luminosity as it convulses into a white dwarf \footnote{For the purposes of this chapter, ``Sun-like" stars are defined to be any stars which will become white dwarfs; significantly more massive stars have different fates.}. This violent transition is known in the astronomy community as the {\it giant branch phases} of evolution. The second section  (``Stage 2") then outlines how planetary system components move around in the system after the parent star has become a white dwarf and eventually reach the white dwarf's {\it Roche sphere}. The Roche sphere is where objects break up easily and become observable. The final section of the chapter (``Stage 3") details our knowledge of the compact debris environment within the white dwarf Roche sphere. This environment is the immediate precursor to accretion onto the white dwarf, where chemical autopsies can be performed (see the subsequent chapter).

\section{Stage 1: Traversing the giant branch phases}\label{GBphases}

\subsection{Introduction}

By the time a planetary system is a few tens of Myr old, its constituent planets, moons, asteroids and comets have become fully formed. Later one-off events, hundreds of Myr or several Gyr in the future, may significantly reshape these primordial configurations or bodies. Potential examples of these one-off events in the solar system include the creation of Earth's moon due to a {\rev giant impact \citep{canup2004}} or the reordering of the ice and gas giant planets due to a crossing of a {\rev gravitational resonance \citep{thoetal1999,tsietal2005}}.

Nevertheless, the older a Sun-like star is, the more likely that its planets have settled into a steady state. Further, although smaller reservoirs of bodies, such as collections of asteroids or planetary rings, are consistently being ground down and replenished, the regions in which they are concentrated would not significantly change unless the planets' orbits do.

However, this relative post-formation quiescence does not last forever. Each star's store of fuel is finite. As the different layers of this fuel, in the form of different elements, are depleted, then the star is transitioning between different giant branch phases. During the giant branch phases the quiescence is broken and the planetary system undergoes significant transformations.

The first layer of the stellar fuel is hydrogen, which the Sun has been fusing into helium for the last 4.6 Gyr. The Sun contains enough hydrogen to continue in its current state, known as the {\it main sequence}, for approximately {\rev the next 6.5 Gyr \citep{veras2016b}}. The vast majority of known exoplanet host stars also have many Gyr remaining on the main sequence before transitioning to the giant branches\footnote{{\rev See the NASA Exoplanet Archive at https://exoplanetarchive.ipac.caltech.edu/.}}.

Consequently, understanding if and how planetary systems would change significantly during this substantial remaining time on the main sequence is important \citep{davetal2014}. Within the solar system, the four giant planets will remain in their current stable orbits until the end of the Sun's main sequence \citep{dunlis1998,veras2016b,zinetal2020} except in the highly unlikely case that a different star flies close enough to these planets to create a significant perturbation \citep{brorei2022}.

The evolution of Mercury, Venus, Earth and Mars, is, however, not as straightforward. Primarily due to Mercury's large eccentricity relative to the other planets, \cite{lasgas2009} found in about 1\% of their simulations, Mercury will collide with Venus or the Sun, destabilizing the rest of the inner solar system. This 1\% value has been subsequently scrutinized. However, because the inner solar system is mathematically chaotic, predicting its future is impossible, even with much more accurate ephemerides \citep{zeebe2015,abbetal2021,abbetal2023,moglas2021,hoaetal2022,brorei2023,mogetal2023}. Nevertheless, the community consensus is that the solar system's four inner planets will very likely survive until the end of the Sun's main sequence lifetime.

This consensus is not surprising, especially given all of the exoplanets that are currently known to orbit stars that have already left the main sequence. Observational limitations restrict our planet discoveries around giant branch stars to giant planets only (as opposed to terrestrial planets). Nevertheless, well over 100 of these giant planets have been discovered \citep{refetal2015,hubetal2019,luhetal2019,nieetal2021,gruetal2019,gruetal2023}, and some giant branch stars exhibit evidence of having recently accreted planets \citep{adaetal2012,steetal2020,sevetal2022}. Further, the discovery of planetary debris disks around these types of stars \citep{bonetal2013,bonetal2014,lovetal2022} indicates the survival of exo-Kuiper belts, and in particular exo-asteroids and the dust they produce from mutual collisions.

This enticing observational evidence motivates exploration of how planetary systems are transformed when their host stars leave the main sequence, which is the subject of the remainder of this section.

\subsection{Orbital shifts}

When stars leave the main sequence, they begin shedding their mass through winds. These winds carry the mass completely out of the system, beyond the star's gravitational influence. As a result, the gravitational potential in the system changes, affecting the orbit of every planet, asteroid, comet, boulder and speck of dust. The effect on the planetary system is significant because the star ends up losing at least 50\% of its mass, and up to 80\%, through {\rev these winds \citep{huretal2000}}.

Further, the loss of mass is not constant because the giant branch phases consist of multiple phases. The two most important of these phases are the {\it red giant phase} and the {\it asymptotic giant branch phase}. The timescales of and mass lost during each phase are a strong function of the stellar mass. 

\begin{figure}
%{\LARGE Fate of Solar System Planets}
\centerline{}
\centering
\includegraphics[width=0.8\textwidth]{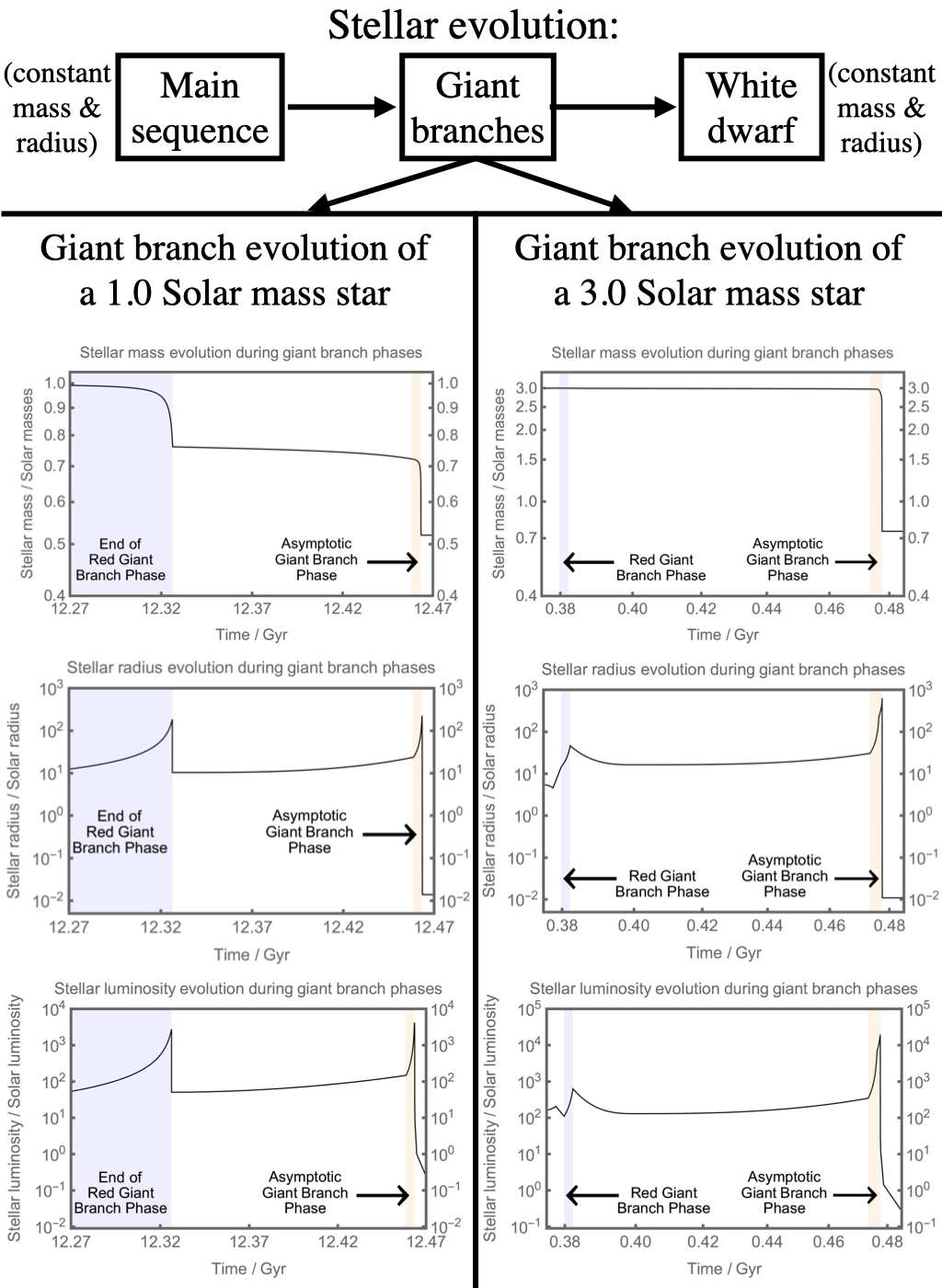}
\caption{Representative mass, radius and luminosity evolution profiles of both Sun-like stars and stars which are three times as massive during the giant branch phases. The stellar mass and radius only change appreciably during the red giant branch and asymptotic giant branch phases, and massive stars reach the giant branch phases more quickly than Sun-like stars. Data for these plots was computed from \cite{huretal2000}.}
\label{Evo}
\end{figure}

Figure \ref{Evo} provides an explicit example. For a star with the same mass and metallicity as the Sun, the red giant and asymptotic giant branch phases last for, respectively, about 700 Myr and 5 Myr, with equal mass loss fractions of approximately 25\%, although this percentage can vary significantly depenending on the model parameters chosen. Instead, for a star whose mass is three times greater, the duration of these phases are about 2.4 Myr and 4.2 Myr, with respective mass loss fractions of approximately 0.019\% and 75\%.

The net result of a star losing mass is the expansion of a body's orbit. A rule-of-thumb is that an orbit typically expands by a factor equal to the ratio of the initial to final mass of the star. If the mass loss rate is particularly large, or a body such as a comet is particularly far away, or the star loses mass very asymmetrically, then the orbit can also stretch, perhaps to a breaking point when the body escapes the system \citep{omarov1962,hadjidemetriou1963,veretal2011,verwya2012,veretal2013a,adaetal2013,veretal2014d,doskal2016a,doskal2016b,regetal2022}. 

However, the orbital changes to a single body become much more interesting in the context of the consequences for other bodies in the system. Although the orbit expansion factor is the same for multiple bodies, the change in gravitational potential alters the conditions in which they can remain stable: because the planet-to-star mass ratio increases, both planet-planet and planet-asteroid interactions become stronger \citep{debsig2002,veretal2013b,voyetal2013}. As a result, multiple planets whose orbits were far enough apart on the main sequence to remain stable during that stellar phase might not remain stable after their parent star starts traversing the giant branches. This phenomenon is not limited to just planets: a planet and an asteroid, for example, can experience this same ``late" instability.

This situation becomes more complex when a companion star is involved, in a so-called {\it binary star system}. Planets, asteroids and comets can orbit one or both of the stars; the former case is referred to as {\it circumstellar} and the latter as {\it circumbinary}. In these binary star planetary systems, both stars may or may not traverse the giant branches concurrently. Mass loss from the system forces the mutual orbit of the stars and any planetary material to all expand their orbits appropriately \citep{kraper2012,portegieszwart2013,kosetal2016}, except if the stars are close to each other (5-10 au). In this case, the stars may form a {\it common envelope} or trigger one or two supernovae, depending on the mass of the stars, leading to much greater mass loss rates than in the single star case and potential ejection of circumbinary material \citep{vertou2012}.

How planetary orbits shift due to stellar mass loss determines their final configurations and is hence a crucial consideration in late-stage population synthesis studies of the currently known exoplanet population \citep{andpop2021,maletal2020a,maletal2020b,maletal2021,maletal2022}. Part of this synthesis includes understanding which systems remain stable and which ones do not. The fates of notable individual systems such as ones with multiple giant planets like HR 8799 \citep{verhin2021} or multiple stars like the circumbinary systems found with the {\it Kepler} space telescope \citep{kosetal2016} or the triple star system HD 131399 \citep{veretal2017c} have received dedicated studies because of their complexity. 

Finally, orbital shifts are not generated solely by stellar mass loss. Another feature of stars which ascend the giant branch phases is that their luminosity increases by up to a factor of about $10^4$. This drastic luminosity increase has two important consequences. The first is that dust is drawn into the star quickly due to an effect known as {\it Poynting-Robertson drag} \citep{bonwya2010,donetal2010,zotver2020}. The second is that pebbles, cobbles, boulders and asteroids may be propelled either inwards or outwards due to an effect known as the {\it Yarkovsky effect} \citep{veretal2015c,veretal2019b}. Overall, orbital shifts from both radiation and gravity would need to be incorporated concurrently in self-consistent treatments of these evolved planetary systems, a prospect which has become more feasible with advances in $N$-body numerical codes \citep{baretal2022,feretal2022}.

\begin{figure}
{\LARGE Fate of Solar System Planets}
\centerline{}
\centering
\includegraphics[width=0.8\textwidth]{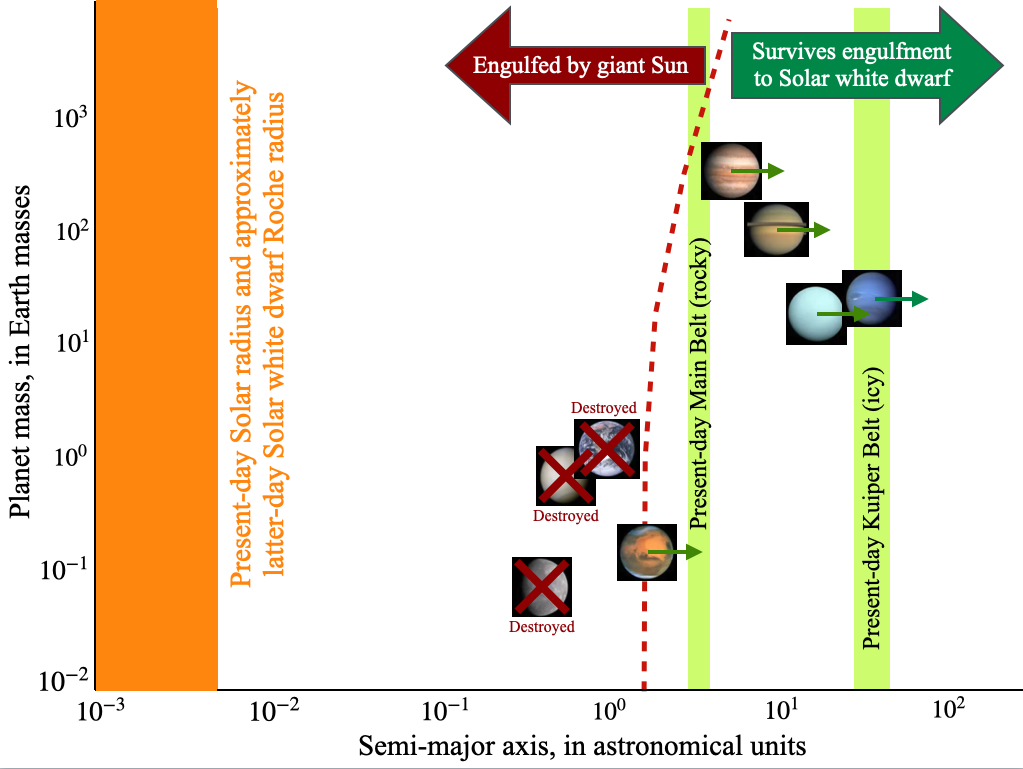}
\caption{The fate of the Solar System's eight planets after the Sun leaves the main sequence in about 6.5 Gyr from now and traverses the giant branch phases. The Sun's radius will increase by a factor of hundreds, enveloping Mercury, Venus and probably Earth. However, Mars, Jupiter, Saturn, Uranus and Neptune will all survive, and be pushed away from the Sun by the amount indicated by the dark green arrows due to the Sun shedding about half of its mass through winds. Although both the asteroid Main Belt and Kuiper Belt (vertical light green strips) will survive engulfment, they will be affected in other ways; due to the Sun's increased luminosity, the Main Belt will be largely ground down and the Kuiper belt will disperse and spread apart. After the Sun becomes a white dwarf, its Roche radius will coincidentally be at a similar distance to its present-day radius ($1R_{\odot}=5\times 10^{-3}$~au). The solar white dwarf's actual photospheric radius will be approximately $5\times 10^{-5}$~au.}
\label{SSFate}
\end{figure}

\subsection{Physical survival}

We have so far described how a star's mass and luminosity changes during the giant branch phases of stellar evolution, and how these changes affect orbital evolution. Another, potentially more destructive aspect of giant branch stars, is their great size. A star's radius can increase by a factor of hundreds while traversing the giant branches, reaching out to a distance of one or more au (one au is equivalent to 215 Solar radii, or about $6.96 \times 10^5$~km).

The consequences for orbiting planets are significant. Consider first the solar system. Its predicted fate is illustrated in Fig. \ref{SSFate}. The Sun's outer envelope will extend far enough to engulf Mercury and Venus, and probably the Earth \citep{goldstein1987,sacetal1993,rybden2001,schsmi2008,iorio2010}. Mars and the outer planets, however, will survive, and approximately double their current semi-major axes due to the Sun losing about half of its mass\footnote{The orbital eccentricity of the surviving planets will also increase by at least $10^{-5}-10^{-4}$ due to tidal convective motions in the star \citep{lanetal2023}.}.

This increase in the star's size occurs while it is losing mass partly because envelope material is more weakly bound at such large radii. Hence, there is a competition between a planetary orbit expanding and the star's radius expanding. In a sense, the planet is trying to outrun the star. Complicating this picture further is the important concept of tidal interactions between the star and planet, and specifically the tidal distortion of the star induced by the planet. This distortion draws the planet closer to the star as the surfaces of the star and planet approach each other. The strength of the tidal interactions is strongly dependent on the mass and radius of the planet and the star, which is why the dashed curve in Fig. \ref{SSFate} bends to the right for higher planetary masses.

Understanding the interplay between tidal interactions and mass loss has led many researchers to compute critical engulfment distances for planetary bodies orbiting stars which traverse the giant branches \citep{villiv2009,kunetal2011,musvil2012,adablo2013,norspi2013,viletal2014,madetal2016,ronetal2020}. The need for so many studies arose because tidal theory is poorly constrained and dependent on a large number of parameters. One of these parameters is stellar mass. Overall, a rough rule-of-thumb is that for each extra solar mass, the critical engulfment distance increases by {\rev another au \citep{musvil2012}}.

The majority of all currently known exoplanets reside within the critical engulfment distance, even when accounting for variations in this distance depending on the tidal theory used. This fraction is not necessarily representative of the true demographics of exoplanets, but rather may reflect our observational bias for more easily finding planets on compact orbits. Many studies about the discovery or dynamics of these compact systems now include estimates for when these planets would be engulfed \citep[e.g.][]{lietal2014,jiaetal2020}, and can highlight peculiar evolutionary histories of the parent star \citep{cametal2019,honetal2023}.

When a star's outer envelope engulfs a planet, the destruction is not instantaneous \citep{staetal2016,jiaspr2018,macetal2018,ocoetal2023a}. In fact, if the planet is large enough, it may survive and expel the envelope before being destroyed \citep{neltau1998,beaetal2011,pasetal2012,beaetal2021,chaetal2021,lagetal2021,meretal2021,yaretal2023}, or expel the envelope while fragmenting into smaller planets \citep{beasok2012}. The critical planet mass above which survival is possible is roughly one Jupiter mass, although survival is likely only for those planets much larger than a Jupiter mass. Planets which do survive this so-called {\it common envelope} evolution will have a net inwards motion due to drag from the envelope, overpowering any orbital expansion from stellar mass loss. Common envelope evolution remains highly uncertain and represents an active area of {\rev research \citep{ropdem2023}}.

Because of the high luminosity of a giant branch star, the atmospheres of surviving planets which are sufficiently close to the star are in danger of partial evaporation \citep{villiv2007,beasok2011,ganetal2019,schetal2019}. Surviving terrestrial planets might have their entire atmospheres evaporated, potentially with implications for the prospects of habitability on those planets \citep{ramkal2016,kozkal2019} and for determining if a new atmosphere is generated on detected planets orbiting stars which have left the main sequence \citep{linetal2022}. An eroding atmosphere, coupled with heating of the surface from increased irradiation, and the loss of a planetary magnetosphere due to giant branch stellar winds \citep{vervid2021}, poses challenges for life surviving during these phases of stellar evolution.

The combination of high stellar luminosity and strong winds also has several other physically significant effects in planetary systems. One is the creation of {\it planetary nebulae}, a term which does not necessarily actually refer to planets, but instead describes a structure of glowing gas around a giant branch star.   The shape of these planetary nebulae may be partially due to the presence of planets interacting with the stellar wind \citep{soker1996,demsok2011,sabsok2018,hegetal2020}, but the origin of the gas itself may be due to sublimation of comets \citep{steetal1990,jura2004,suetal2007,maretal2022}.

Dust belts and disks are subject to dust destruction, blowout and collisional evolution, while coupled with orbital expansion from stellar mass loss \citep{bonwya2010,donetal2010,veretal2015c,zotver2020}. Even large asteroids may be broken apart from spinning too quickly (named the YORP effect), all entirely due to high stellar luminosity \citep{veretal2014c,versch2020}, perhaps after changing shape \citep{katz2018}, and maybe followed by inward or outward drift from the Yarkovsky effect \citep{veretal2015c,veretal2019b}.

\begin{figure}
%{\LARGE Potential Pollutants of White Dwarfs}
\centerline{}
\centering
\includegraphics[width=1.0\textwidth]{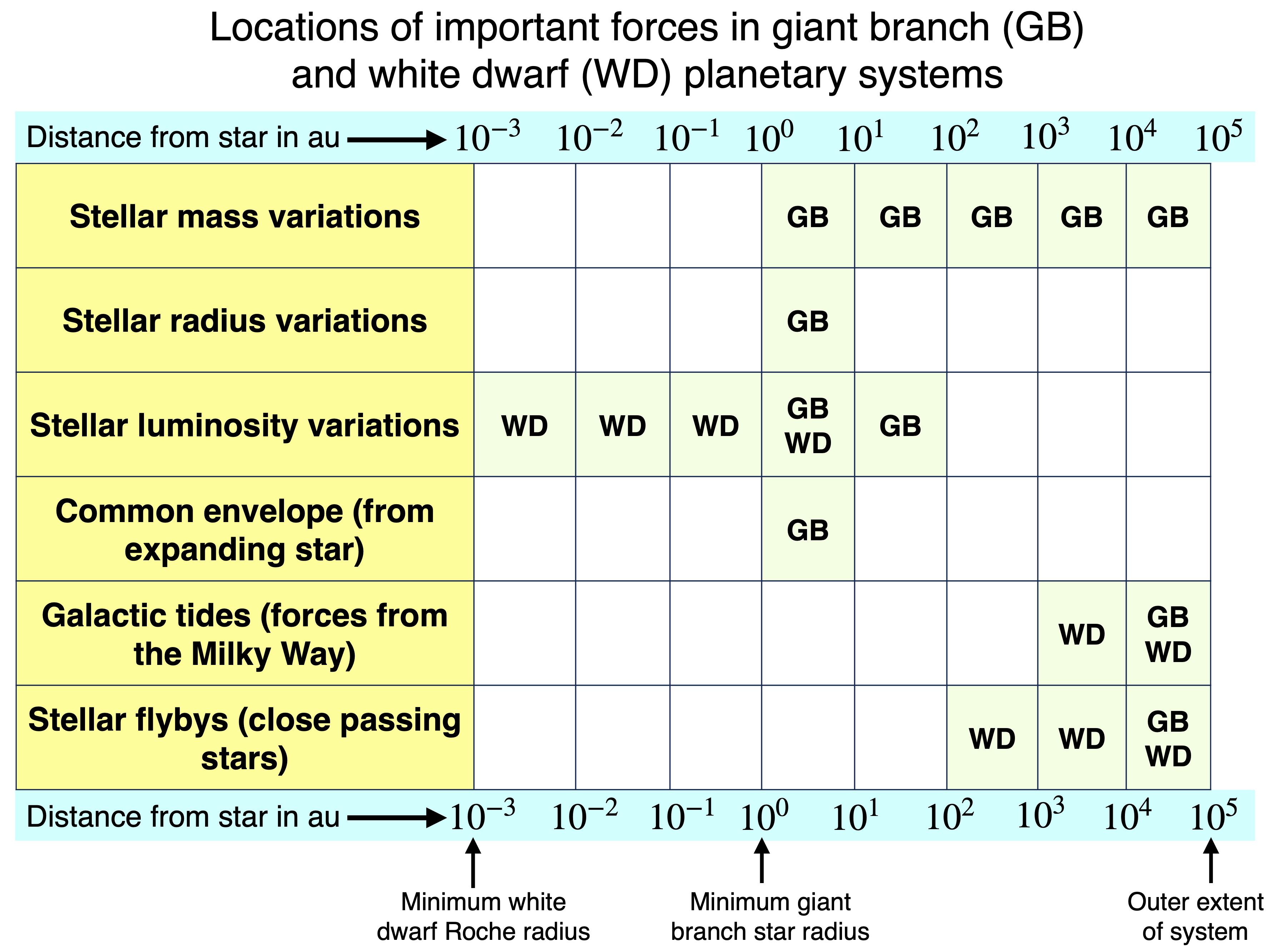}
\caption{Order-of-magnitude distance estimates of which forces act upon planetary systems after the parent star has left the main sequence. {\rev The first four forces are internal to the planetary systems, whereas the last two, from the cumulative effect of all stars in the Milky Way and from individual passing stars, are external.} Unlike for giant branch stars, white dwarf stars have constant masses and radii. Also, giant branch lifetimes are relatively short, allowing Galactic tides and stellar flybys to penetrate more deeply in white dwarf planetary systems.}
\label{Forces}
\end{figure}

The resulting picture is one where extant minor planets or dust may be significantly redistributed within the system during the giant branch phases. Planets which survive stellar engulfment may then exist in a sea of smaller particles, rather than being located at a specific distance from more concentrated reservoirs of smaller particles, as on the main sequence. Figure \ref{Forces} provides a summary of the different forces involved in determining where and how planetary constituents can survive the giant branch phases. 

\subsection{Compositional alteration}
 
Planetary bodies which do survive the giant branch phases of stellar evolution may be {\rev compositionally} altered in two important ways. The first is through depletion of volatiles due to the star's high luminosity \citep{jurxu2010,juretal2012,malper2016,malper2017a,malper2017b,katz2018}, leading to largely ``dry" asteroids except perhaps in the deep interiors of the larger members. Observations of chemical species after the giant branch phases, combined with a quantification of volatile depletion, can also help one infer formation properties of particular systems \citep{haretal2021}.

The second way is through the compositional effect of the stellar wind on planetary atmospheres. Giant planet atmospheres can accrete enough of the stellar wind to change their C/O, O/H, H$_2$O, and CH$_4$ abundances by a potentially observable amount \citep{spimad2012}. For terrestrial planets, the change in mixing ratios must be treated in concert with the extent of the evaporation of the atmospheres \citep{kozkal2019,linetal2022}.
 
The consequences of these {\rev compositional} alterations, as well as the general survival of planetary bodies during the giant branch phases of evolution, are {\rev observed in} older systems, which is the subject of the remainder of this chapter.

\section{Stage 2: Delivering material to the white dwarf's Roche sphere}\label{toWD}

\subsection{Introduction}

At the conclusion of a star's giant branch phases, a critical point is reached with the star's remaining fuel, which would be composed of primarily C and O. The star is no longer hot enough to fuse such heavy elements. At this point, these elements build up into an inert core, and all of the outer layers of the star are shed. What remains is a burnt core known as a {\it white dwarf}.

A white dwarf's radius is relatively tiny; when the Sun becomes a white dwarf, its radius will shrink to about $5 \times 10^{-5}$~au (1.2 Earth radii, or 7500 km). However, throughout this phase, the Sun will lose only about half of its mass. Therefore, its surface gravity will be six orders of magnitude higher than the current surface gravity of the Sun. This sudden increase in density will drag down any elements in the star's photosphere that are heavier than {\rev H or He \citep{paqetal1986,koester2009}}.

Hence, white dwarfs have almost pristine atmospheres composed of either H and/or He. This concept becomes very important for planetary science because it enables chemical autopsies to be performed on any planetary material which accretes onto the white dwarf atmosphere: we know that any elements in a white dwarf atmosphere heavier than He must come from accreted planetary material, with few exceptions (for instance, the star's radiation can sometimes levitate certain atomic species {\rev for the youngest white dwarfs; \citealt*{chaetal1995}}). 

Further, and crucially, we know that accreted planetary material is a very common phenomenon. In fact, dedicated surveys of different patches of the Galactic neighborhood of the Sun reveal that one in every two to four white dwarfs contain accreted planetary material \citep{zucetal2003,zucetal2010,koeetal2014}.

The subsequent chapter will cover the exciting chemical trends and links with composition, differentiation and formation that we observe. The remainder of this chapter will instead focus on how all of the surviving planetary material from the giant branch phases of evolution reaches the tiny white dwarf. In fact, reaching the white dwarf's atmosphere is not requisite; planetary material needs only to reach the white dwarf's {\it Roche sphere radius} (or {\it Roche radius}), where material will fragment due to strong tidal forces, subsequently evolve and then eventually accrete onto the photosphere. The approximate location of a white dwarf's Roche radius is, coincidentally, at about $1R_{\odot}$, or hundreds of white dwarf radii (see Fig. \ref{SSFate}).

This section will describe how planetary material might be delivered down to $1R_{\odot}$ from au-scale distances after the star has become a white dwarf, and the next section will describe the complex debris environment that we observe around and within $1R_{\odot}$ of white dwarfs. First we describe our knowledge from observations.

\subsection{Known planetary material exterior to the white dwarf Roche sphere}

Around giant branch stars, planets and disks are primarily observed soon after the star has left the main sequence, on the red giant branch. Observations of planets or disks around asymptotic giant branch stars are rarer, and only one unconfirmed planet candidate actually exists \citep{keretal2016} in this final phase before the star technically becomes a white dwarf. Unfortunately, due to observational biases, we have little information on the architectures and frequencies of wide-orbit planets that can survive stellar evolution.

The transition to a white dwarf is not necessarily sharp \citep{soker2008}, particularly when the star is still very hot and luminous, even after its outer layers have mostly dissipated. In this transition phase, observations remain rare but important. The Helix planetary nebula was discovered in this transition phase, with dust detected at tens of au \citep{suetal2007}. Although the planetary origin of this dust is still debated \citep{claetal2014,maretal2022}, its presence provides observational evidence of planetary material that might just be starting to make its journey to the parent white dwarf's Roche radius \citep{verhen2020}.

Newly formed white dwarfs have temperatures of $10^5$~K, but cool quickly. This temperature is halved after just 10~Myr, and reduced to $10^4$~K after 1~Gyr. The age of a star after it has become a white dwarf is known as its {\it cooling age}. Young white dwarfs are luminous enough to easily evaporate atmospheres of nearby planets \citep{schetal2019}: white dwarf WD~J0914+1914, with a cooling age of just 13.3 Myr, is evaporating an ice giant planet at a separation of just 0.07~au \citep{ganetal2019,verful2020,veras2020,zotetal2020}.

The other known planets orbiting white dwarfs are all gas giants and orbit much older white dwarfs, all with cooling ages over 1~Gyr old. These planets include WD~0806-661~b, at a planet-star separation of about 2500~au \citep{luhetal2011,rodetal2011}, PSR~B1620-26~(AB)~b, at a planet-star separation of 23 au \citep{thoetal1993,sigetal2003,beeetal2004,sigtho2005}, MOA-2010-BLG-477L~b, at a planet-star separation of a few au \citep{blaetal2021}, and WD~1856+534~b, at a planet-star separation of just 0.02~au \citep{munpet2020,vanetal2020,aloetal2021,meretal2021,xuetal2021}.

These 5 planets orbiting white dwarfs span 5 orders of magnitude in star-planet separation (0.02-2500~au), which already reveals to us that their dynamical origins are diverse. MOA-2010-BLG-477L~b's location at a few au might have just exceeded its progenitor star's critical engulfment distance, and the planet has since remained undisturbed. WD~0806-661~b (at 2500~au) was either gravitationally scattered outward, or represents a captured free-floating planet. Both WD~J0914+1914~b (0.07~au) and WD~1856+534~b (0.02~au) must have been either dragged inwards in their progenitor's envelope -- and survived the process -- or were more suddenly scattered inwards, and then dynamically settled due to tidal influence with the star. This variety of possibilities has prompted theoretical investigations about planet-white dwarf interactions \citep{verful2019,veretal2019a,ocolai2020} as well as the possibility of forming new planets only after the star has become a white dwarf \citep{beasok2014,schdre2014,voletal2014,beasok2015,hogetal2018,ledetal2023},

Further, each of the 5 known planets (WD~J0914+1914~b, PSR~B1620-26~(AB)~b, MOA-2010-BLG-477L~b, WD~1856+534~b, WD~0806-661~b) was discovered with a completely different technique: photometry, spectroscopy, imaging, microlensing, and pulsar timing. This variety of discovery methods, as well as the enticing prospect of understanding this planet population better, has prompted a large number of recent detection estimates and efforts 
\citep{faeetal2011,xuetal2015,sanetal2016,sheetal2016,vanvan2018,corkip2019,dametal2019,danetal2019,tamdan2019,verwol2019,krzetal2020,braetal2021,kanetal2021,moretal2021,vangrootel2021,waletal2021,lucetal2022,coletal2023,kubetal2023}.

Two instruments of note which are expected to make major advances in the known population of planets orbiting white dwarfs are {\it Gaia} and the {\it James Webb Space Telescope} (JWST). Gaia is expected to discover multiple planets through yet another technique, astrometry \citep{peretal2014,sanetal2022}. JWST might also discover planets and help chemically characterise their atmospheres in ways which have not been possible before \citep{kozetal2018,kaletal2020,kozetal2020,limetal2022}.

Unlike WD~J0914+1914~b, most of the known planets orbiting white dwarfs do not expel their atmospheres onto the white dwarf. Nevertheless, their positions and masses play a crucial dynamical role in shepherding smaller material towards the white dwarf. This material is what eventually enters the white dwarf's Roche radius, breaks up, and accretes onto the white dwarf's atmosphere, where the chemical autopsy can be performed best.

\begin{figure}
{\LARGE Potential Pollutants of White Dwarfs}
\centerline{}
\centering
\includegraphics[width=1.0\textwidth]{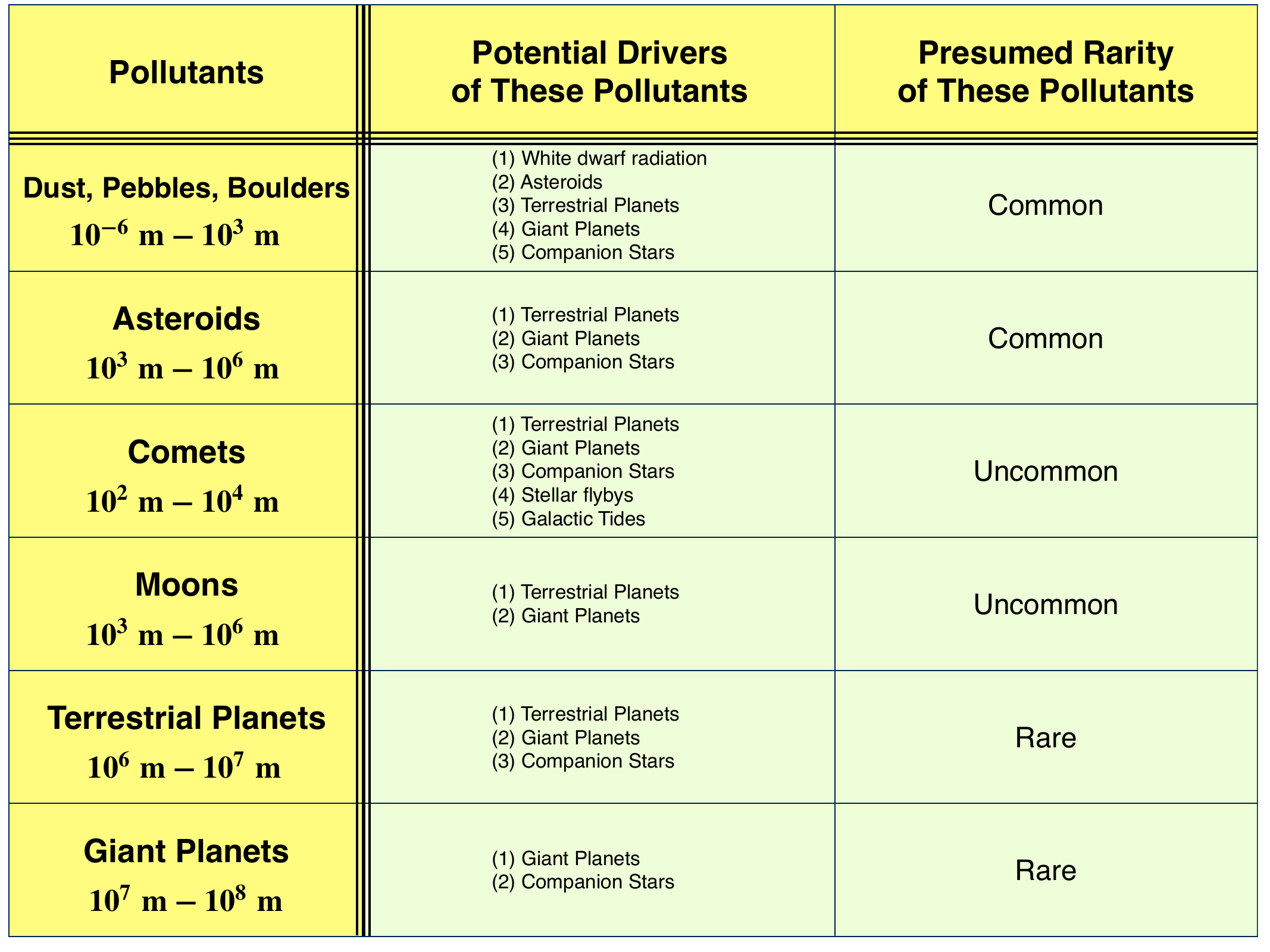}
\caption{Summary of the different parts of planetary systems which can be transported to the Roche sphere of a white dwarf, eventually {\it polluting} it. The numbers in the first column refer to the typical radii of the corresponding objects. Listed in the second column are the different mechanisms or objects which drive this transport, sometimes in concert with one another.}
\label{PolTable}
\end{figure}

\begin{figure}
%{\LARGE dssd}
\centerline{}
\centering
\includegraphics[width=1.0\textwidth]{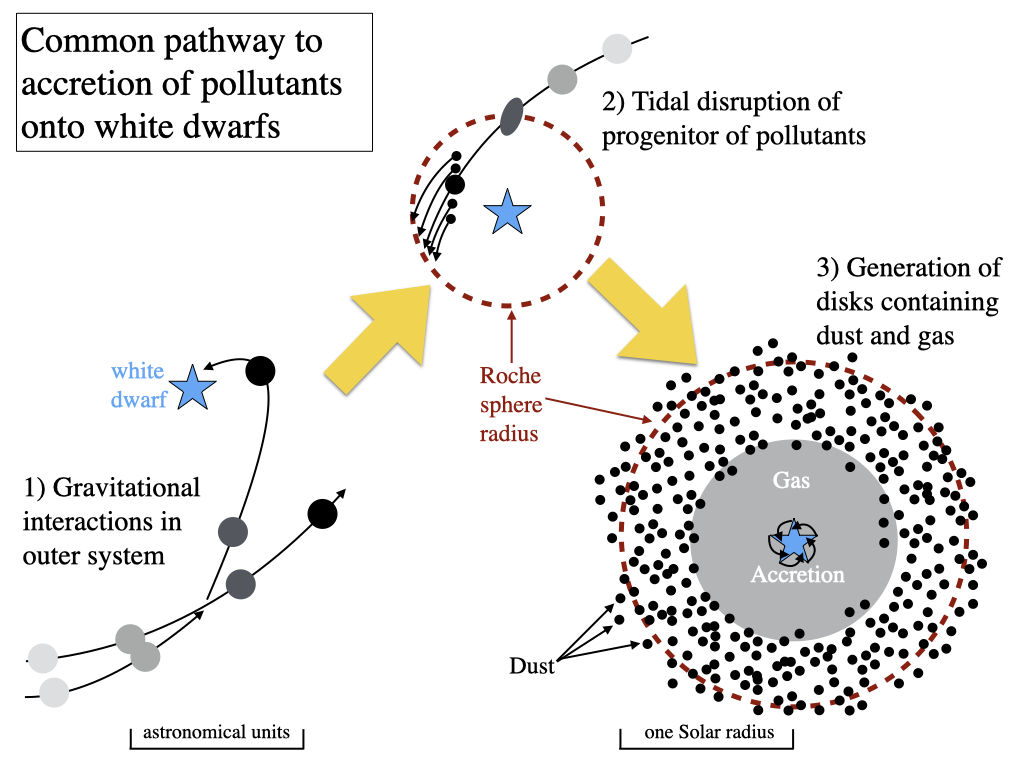}
\caption{An assumed-to-be common method of transport of planetary material to the Roche sphere of the white dwarf, where breakup into disks of dust and gas occurs. Close to the white dwarf, dust becomes sublimated and only gas is present. This gas is accreted onto the white dwarf, polluting its atmosphere and providing us with bulk chemical constraints on the exoplanetary material. }
\label{Pathway}
\end{figure}

Figures \ref{PolTable} and \ref{Pathway} make this notion concrete. The table in Fig. \ref{PolTable} uses the very common community term {\it pollutant} to indicate in the first column the type of material that enters the Roche radius. The second column then outlines the different classes of objects which can drive the pollutants to the Roche radius. The schematic in Fig. \ref{Pathway} illustrates a particularly common pathway -- gravitational scattering -- through which objects can approach a white dwarf's Roche radius and be broken up into dust and gas.

The remaining subsections will describe each potential polluter in turn. Some are more likely than others based on theoretical considerations and our observational knowledge of the material within the Roche radius (next section). As this subsection has demonstrated, however, our current observational knowledge of material outside of the white dwarf's Roche radius is currently limited to effectively just 5 planets and the Helix nebula.

\subsection{Delivery mechanisms: dust, pebbles and boulders as polluters}

Dust, pebbles and boulders represent the only class of object which may be drawn towards the white dwarf's Roche sphere by radiation or magnetism alone \citep{veretal2015b,veras2020,zhaetal2021,veretal2022}; see Figure \ref{PolTable}. Gravitational interactions from larger objects may help, but are not necessary.

In this respect, provided that a sufficient amount of dust exists after the giant branch phases of stellar evolution, it can represent the primary polluter in some systems. These dust, pebbles and boulders may be generated from the break-up of asteroids or planets at any time, or even from smaller scale crater impacts on those bodies \citep{verkur2020}.

The timescale for radiation or magnetism to drag in dust, pebbles and boulders depends strongly on their size and the magnitude of the luminosity of the white dwarf, which relates to its cooling age. Another dependence is on the distance to the white dwarf; closer-in dust is drawn in more quickly. Hence, the breakup of minor planets at the white dwarf's Roche radius can produce fragments that reside just outside of that boundary; these are quickly enveloped \citep{lietal2021}. At such close distances, these fragments may also collide with {\rev another}, breaking into even smaller particles which then may be dragged in more quickly \citep{malamud2021,broetal2022,broetal2023a,broetal2023b}.  

In binary systems where the binary companion to a white dwarf is emitting winds, dust may originate from this companion and reach the white dwarf's Roche radius \citep{perets2011,perken2013}. The amount of this dust is non-negligible when the binary separation is within about 1~au \citep{debes2006,veretal2018b}. For wider separations, observed pollutants must arise from a planetary system rather than from the winds of the binary star.

\subsection{Delivery mechanisms: asteroids as polluters}

Asteroids, or perhaps more generally, dry minor planets, represent the population of bodies which is most often cited as the primary polluters of white dwarfs \citep{jura2003}. This representation has become canonical over the decades due to both strong dynamical and chemical arguments. For the latter, the subsequent chapter will demonstrate that the chemical compositions seen in white dwarf atmospheres are even more diverse than what we see amongst the asteroid and meteorite families in the solar system \citep{putxu2021}.

From a dynamical point-of-view, asteroids are effectively massless compared to planets. As a result, a single surviving planet can easily perturb asteroids gravitationally and excite their orbital eccentricities to high enough values to reach the white dwarf Roche sphere \citep{bonetal2011,debetal2012,frehan2014,picetal2017,antver2019,veretal2021,jinetal2023,mcdver2023,veretal2023a}. Although not all orbital configurations will allow for this eccentricity excitation \citep{antver2016,veretal2020b}, adding additional planets or companion stars increases the parameter space for asteroid pollution \citep{bonver2015,hampor2016,petmun2017,steetal2017,musetal2018,smaetal2018,smaetal2021,ocoetal2022,stoetal2022}.

Other dynamical lines of evidence support asteroids as likely polluters. The first is that in our solar system, the pollution of the solar white dwarf will arise almost entirely from minor planets which survive break-up from the Sun's giant branch luminosity \citep{lietal2022}. The second is that because a planetary system can host thousands of minor planets, the opportunities to pollute are numerous, potentially commensurate with the regularity seen in survey observations of populations of white dwarfs \citep{zucetal2003,zucetal2010,koeetal2014}.

Asteroids which can generate pollution do, however, require the presence of larger bodies in the system to act as perturbers. Asteroids themselves are not massive enough to gravitationally perturb one another into a white dwarf's Roche sphere, but instead require pollution drivers at least as massive as that of Earth's moon \citep{verros2023}. These larger objects need first to be fully formed and then would need to survive the giant branch phases of evolution. On the other end of the mass spectrum, there is no upper limit to the size of perturbers, meaning that no planets need to be present if a binary star can act as an efficient pollution driver \citep{hampor2016}.

\subsection{Delivery mechanisms: comets as polluters}

Comets, or volatile-rich minor planets, have long been disfavoured as white dwarf polluters based on chemical grounds (see the subsequent chapter). However, repeatedly this judgement has {\rev been} challenged. Observations of water-rich signatures in white dwarf atmospheres \citep{faretal2013,radetal2015,genetal2017,xuetal2017,hosetal2020} now strongly suggest that comets occasionally pollute some white dwarfs. Although comets are still not assumed to be the primary pollutants, they provide some of the most interesting exceptions.

Unlike dry asteroids, volatile-rich comets can originate anywhere from distances of tens of au to hundreds of thousands of au away from their parent star (as well as at intermediate distances; \citealt*{rayarm2013}). This wider range of locations allow for more delivery mechanisms than what is available for asteroids. Two in particular include influences from outside of the parent planetary system: stellar flybys and Galactic tides \citep{alcetal1986,paralc1998,veretal2014a}; see Figure \ref{Forces}. The most distant comets are also strongly affected by changes to the motion of the white dwarf in the Galaxy when the white dwarf is born \citep{stoetal2015}. The presence of binary stellar companions \citep{steetal2017} and planets \citep{caihey2017} can help comet pollution prospects, but {\rev planets have also been shown to hinder comet pollution depending on the system architecture \citep{ocoetal2023b}}.

In the solar system, comets lose material to sublimation during each close passage of the Sun, partially altering their orbits due to conservation of linear momentum. Although this effect would also hold in white dwarf planetary systems, the orbital changes are unlikely to drive the comet into the white dwarf's Roche sphere \citep{veretal2015a} without some other perturbing influence.

\subsection{Delivery mechanisms: moons as polluters}

The prospects for moons to represent potential polluters of white dwarfs primarily rely on dynamical rather than chemical arguments because chemically distinguishing asteroids from moons may be difficult. In fact, in only one case has the existence of an exo-moon been inferred based on the composition of the material accreted by a white dwarf, notably its Be content \citep{doyetal2021}.

In terms of the mass budget, in our solar system moons contain about two orders of mass more than asteroids. In terms of sheer numbers, there are dozens of moderate sized moons, which represents orders of magnitude less than the number of asteroids, but one order of magnitude more than the number of planets. The prevalence of white dwarf pollution suggests that if massive bodies are primarily responsible, then they need to be broken up in such a way that delivery can persist for a long time, and not be delivered all at once.

In order for a moon to reach the white dwarf's Roche sphere, either its parent planet has to be perturbed there (an unlikely possibility, as described in the subsequent subsections) or the moon must be stripped from its parent planet and then perturbed -- as a minor planet orbiting the star in its own right -- towards the white dwarf. The stripping can occur either from the moon slowly migrating beyond the planet's gravitational influence from tidal interactions, or more violently due to a gravitational scattering event. The latter scenario has {\rev been explored \citep{payetal2016,payetal2017,maretal2019,trietal2022}} and been found to be effective in a non-negligible number of cases. Overall, pollution from moons is less likely than from asteroids, but more likely than from planets.

\subsection{Delivery mechanisms: terrestrial planets as polluters}

Observations suggest that the frequency of pollution at white dwarfs \citep{zucetal2003,zucetal2010,koeetal2014} is much higher than what could be explained if all the pollution was due to planets. Pollution signatures at most persist for Myr, orders of magnitude smaller than a typical $\sim$Gyr-old white dwarf. There are simply not enough planets per system to generate all those signatures. 

Nevertheless, many investigations have modelled the evolution of terrestrial planets scattering off of one another and off of giant planets as a star is transformed into a white dwarf \citep{debsig2002,vergan2015,hampor2016,veretal2016a,veras2016b,veretal2018a,maletal2020a,maletal2020b,maletal2021,maletal2022}. They find that terrestrial planets may be scattered to the white dwarf Roche sphere at any cooling age, but often just once per system, if at all. The greater the number of initial planets, the greater the likelihood of these scattering events during the white dwarf phase.

Although these results de-emphasize the importance of terrestrial planets as polluters, they should not diminish the role of terrestrial planets as pollution drivers of smaller objects.

\subsection{Delivery mechanisms: giant planets as polluters}

Giant planets represent even rarer polluters. Effectively, the only two potential pollution drivers in this case are other giant planets or companion stars. As shown in the previous two chapters, giant planets form more infrequently than terrestrial planets. Also, companion stars represent effective scatterers only if they reside sufficiently close to the white dwarf, or can change their orbit to do so.

Regardless, partly because the only detected planets orbiting white dwarfs are currently giant planets, the evolution of giant planets across all phases of {\rev stellar} evolution -- even excluding survival within a giant branch envelope -- remains a well-investigated topic \citep{debsig2002,veretal2013b,musetal2014,vergan2015,veretal2016a,veretal2017b,veretal2018a,steetal2018,munpet2020,maletal2020a,maletal2020b,maletal2021,maletal2022,ocoetal2021}. Further, we know of one important case of a giant planet which actually is polluting its parent white dwarf: the evaporating atmosphere of WD~J0914+1914~b is transporting detectable quantities of H, O and S into the white dwarf Roche sphere \citep{ganetal2019}.

\section{Stage 3: Processing debris within the white dwarf Roche sphere}\label{atWD}

\subsection{Introduction}

Having reached the Roche sphere radius of a white dwarf, what then happens to planetary material? The answer is complex, and one which is greatly enhanced by abundant observations. This section will describe both the observations and theory of this narrow region extending from about $5\times 10^{-5}$~au to $5\times 10^{-3}$~au.

The mechanics of an object breaking up at a Roche sphere, and the aftermath, involves fragments and debris which are formed into structures sometimes known as disks or rings. This result is not surprising because disks and rings represent the lowest energy state that astrophysical systems which are supported by rotation settle into, and can be seen at all scales. The debris around the white dwarf can be composed of both gas and dust, and exist revolving around within the Roche sphere for potentially Myrs before being accreted onto the white dwarf's atmosphere.

\begin{figure}
\centerline{}
\centering
\includegraphics[width=0.8\textwidth]{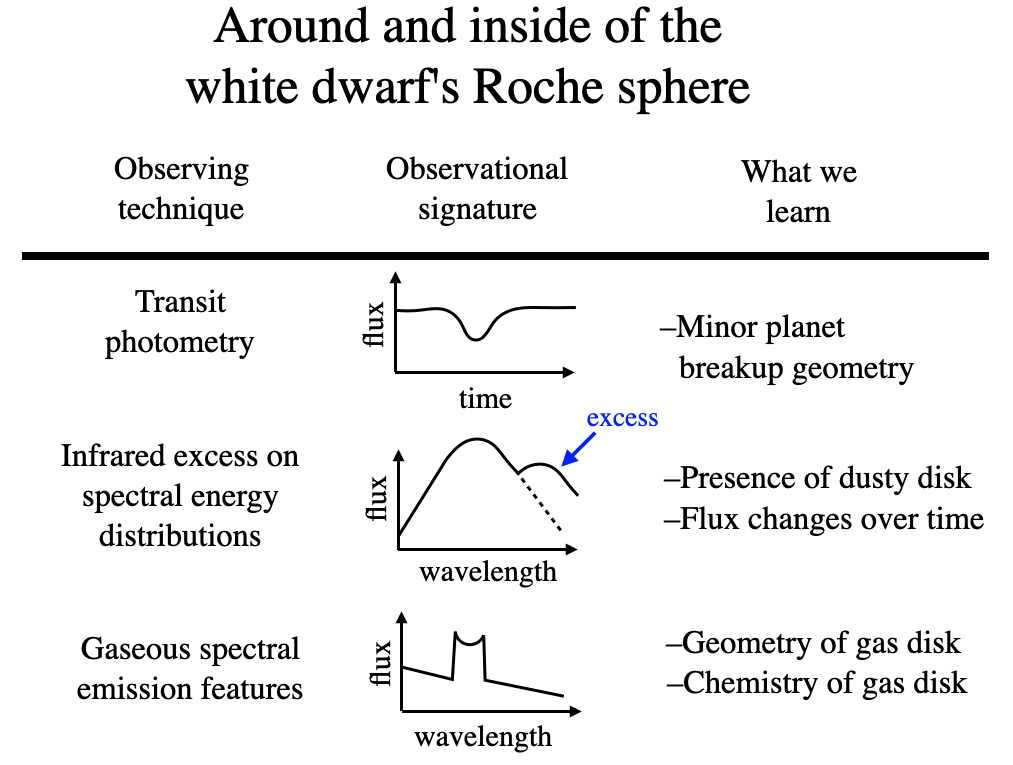}
\caption{The three main types of observations of debris around and within a white dwarf's Roche sphere radius, located at about $1R_{\odot}$, or $5\times 10^{-3}$~au, from the centre of the star. These observations are independent of those of planetary material which is eventually accreted onto the star's photosphere (the subsequent chapter), located at a distance of about $5\times 10^{-5}$~au.}
\label{ObsTable}
\end{figure}

Observations of this debris rely on photometry and spectroscopy, and provide us with different types of physical and chemical constraints. Figure \ref{ObsTable} summarizes the three main types of observations in this region. Each observation type will be described in its own subsection. These observations have helped constrain theoretical models for the formation and evolution of this debris, which will also be described in two dedicated subsections.

\subsection{Observations of debris: transits}

{\it Transit photometry} refers to the dimming of the white dwarf due to material passing in front of it. Because the material orbits the white dwarf, the dimming is periodic, and any change in the amount of dimming per orbital period indicates dynamical activity. Further, a single {\rev orbiting object, such as} a planet, would produce a pronounced a sharp dimming feature. Alternatively, a messy collection of debris, dust and gas would produce a transit curve with lots of peaks and troughs of different extents.

When applied to Sun-like stars, transit photometry has proven to be the most successful exoplanet-hunting technique, being responsible for the discovery of the majority of the currently known exoplanets. Around white dwarfs, so far this technique has been responsible for the discovery of just one planet, WD~1856+534~b, which resides outside of the Roche sphere \citep{vanetal2020}. The reason for this low yield is perhaps because the technique is only sensitive to close-in planets which would be unlikely to survive the star's evolution.

However, the technique has been much more successful at charting the evolution of debris orbiting within the white dwarf's Roche sphere. This debris is commonly inferred to result from the active breakup of a minor planet, which conforms with the common assumption that white dwarfs are polluted by asteroids. Multiple debris clumps can share the same orbit around a white dwarf in a gravitationally stable fashion, even near the Roche sphere radius \citep{veretal2016b}. In few other contexts within exoplanetary science can an individual exo-minor planet be analyzed and tracked. 

Over 8 white dwarf systems now have transit detections of planetary debris \citep{guietal2021}. The first of these discoveries, for the WD~1145+017 system, was seminal \citep{vanetal2015}. That discovery illustrated starkly the important link between the previously theorized survival of planetary material across stellar evolution (at the time the only planets known around white dwarfs were the distant objects PSR~B1620-26~(AB)~b and WD~0806-661~b) and the observed signatures of metal pollution in white dwarf atmospheres. Follow-up observations and theories about the minor planet breaking up around WD~1145+017 were extensive \citep{rapetal2016,zhoetal2016,faretal2017b,guretal2017,haletal2017,veretal2017a,faretal2018a,izqetal2018,xuetal2018b,xuetal2019a,sheetal2019,duvetal2020,ocolai2020,budetal2022}.

Other notable white dwarfs with transiting debris signatures are ZTF~J0139+5245 \citep{vanderboschetal2020}, ZTF~J0328-1219 \citep{vanderboschetal2021} and WD~1054-226 \citep{faretal2022}. The orbital period of the transiting debris orbiting ZTF~J0139+5245 is about 550 times longer than the 4.5-hour orbital period of the debris in WD~1145+017. This striking difference indicates that the debris around ZTF~J0139+5245 is {\rev produced by an extended stream} on a highly eccentric orbit, most of which is actually outside of the Roche sphere \citep{veretal2020a}. In the ZTF~J0328-1219 system, two specific periodicities of 9.9 and 11.2 hours have been identified, representing evidence of two distinct orbiting clumps. The periodicity structure in the WD~1054-226 system is even more complex in the sense that a baseline level of flux has not yet been observed. 

{\rev Overall, these transit signatures showcase an astounding variety of features and parameter ranges which remain difficult to interpret.}

\subsection{Observations of debris: infrared excess}

An entirely different way of observing debris at and within the Roche sphere radius is with {\it spectral energy distributions}. These are plots of spectral flux versus wavelength for different white dwarf systems. The large bump on the plot represents the white dwarf; an accompanying smaller bump (see Fig. \ref{ObsTable}) indicates the presence of dust outside of the white dwarf. This accompanying bump is known as the {\it infrared excess}.

The geometric configuration of this dust cannot be identified unambiguously from the infrared excess. Instead, the mapping between the infrared excess and geometric configuration is degenerate. If one assumes that the dust resides in a flat opaque configuration, then the disk's line-of-sight inclination and its inner and outer boundary can be estimated \citep{jura2003}. This assumption typically yields outer boundaries which are similar to the Roche sphere radius (about $1.0R_{\odot}$), and inner boundaries which are about half of that value (about $0.5R_{\odot}$).

We now know that the assumption of the dust residing in a flat, circular opaque disk is usually poor. \cite{juretal2007a} found that a partially warped disc would better fit the infrared excess around the white dwarf GD~362, {\rev and \cite{juretal2007b} similarly found that the infrared observations of the white dwarf GD~56 cannot be reproduced with a flat disc model. \cite{genetal2021} was not able to use a flat disc model to reproduce  the observations of WD J0846+5703}. \cite{denetal2016} demonstrated that eccentric, rather than circular discs, can in some instances reproduce infrared excesses well. By using an independent radiative transfer-based method of analyzing the polluted white dwarf G29-38, \cite{baletal2022} found that a vertically high, but narrow ($0.1R_{\odot}$), dust ring with a large opening angle is consistent with the observations. This structure has a much greater vertical extent than can be approximated by the canonical flat disk assumption. 

Over 60 white dwarf systems have exhibited clear detections of infrared excess since 1987 \citep{zucbec1987,graetal1990}, including one instance of a circumbinary dusty structure \citep{faretal2017a}. This population is now large enough to motivate reports of multiple discoveries at once and justify {\rev frequency analyses \citep{faretal2009,faretal2010b,baretal2012,beretal2014,rocetal2015,farihi2016,twiletal2019b,cheetal2020,manetal2020,rogetal2020,xuetal2020,laietal2021,wanetal2023}}. The frequency analyses indicate that only a few percent of white dwarfs contain detectable infrared excess, while between 25-50\% are polluted \citep{zucetal2003,zucetal2010,koeetal2014}. This mismatch is likely due to observational bias \citep{bonetal2017} because the theoretical expectation is that dust surrounds nearly every white dwarf which is polluted.

A few white dwarfs with infrared excesses are individually notable. One is G29-38, which is particularly bright and close to the Earth. As a result, it is the only dusty structure which has been chemically probed for silicate minerology \citep{reaetal2005,reaetal2009}. Another notable white dwarf is LSPM J0207+3331, because it has by far the oldest cooling age (about 3 Gyr) of any white dwarf with an infrared excess \citep{debetal2019}.

One exciting feature of white dwarf infrared excesses is the flux variability they demonstrate over yearly timescales. This variability is indicative of dynamical activity \citep{beasok2013}. The most dramatic case so far known is that of WD~0145+234, whose infrared flux changed by one order of magnitude in under a year \citep{wanetal2019,swaetal2021}. Less dramatic, but still notable individual cases have been mounting \citep{xujur2014,xuetal2018a,faretal2018b}, and now flux changes at the tens of percent level per year are assumed to be a ubiquitous feature of white dwarf infrared excesses \citep{swaetal2019b,swaetal2020}.

\subsection{Observations of debris: gas emission lines}

Infrared excesses indicate the presence of only dust, not gas. However, within a white dwarf Roche sphere, gas can be generated by both collisions and sublimation. Collisions may occur anywhere with a sufficient number density of fragments. Sublimation could occur only if a fragment is sufficiently close to a white dwarf; for the hottest and youngest white dwarfs, this boundary exceeds the Roche sphere, whereas for the coolest and oldest, this boundary may be {\rev as low as $0.2R_{\odot}$ \citep{veretal2023b}}.

In all of these cases, a different detection technique is required than to detect dust. Gas orbiting white dwarfs is found using {\it emission line spectroscopy}. As a result, the chemical composition of the gas can be identified immediately, unlike for the dust. This feature helped to confirm the existence of the evaporating planet WD~J0914+1914~b through the identification of H, S and O in the gas surrounding the white dwarf, as well as in the star's photosphere \citep{ganetal2019}.

A more easily detectable chemical element in gas surrounding white dwarfs is Ca, a very common polluter (see the subsequent chapter). In fact, the {\it Ca II triplet feature} (see Fig. \ref{ObsTable}) has become not only a way to identify gas, but also to characterize its geometry. The morphology of this observational signature was instrumental in resolving the degeneracies inherent with infrared excess for identifying the inner and outer edges of orbiting debris, and confirming that these boundaries {\it typically} do reside at distances of about $0.5R_{\odot}$ and $1.0R_{\odot}$ from the centre of the white dwarf \citep{ganetal2006,ganetal2007,ganetal2008}. 

The known white dwarfs containing orbiting gas within the Roche sphere now number over 20 \citep{meletal2010,meletal2020,denetal2020,manetal2020,genetal2021}. One of these is WD~1145+017, the same white dwarf which hosts the first transit photometry detection of minor planet breakup \citep{xuetal2016,redetal2017,foretal2020}. Some gas signatures around white dwarfs have now been the focus of applications of spectral abundance models \citep{haretal2011,haretal2014,steele2021}.  

Like with infrared excesses, gas emission signatures can show variability \citep{wiletal2014,denetal2018}. {\rev In one case, the emission signatures of white dwarf SDSS~J1617+1620 decreased over six years until it fell below the detectability threshold \citep{wiletal2014}. For emission signatures which maintain their strength,} when a sufficient time baseline of observations of this gas is taken, then the knowledge of the structure's geometry can be derived through the use of {\it Doppler tomography}. Doppler tomography allows one to create a velocity map of the gas structure. This map illustrates the intensity pattern of the gas \citep{manetal2016a,manetal2021}. One notable case of variability is with white dwarf SDSS J1228+1040 \citep{manetal2019}. This variability indicates the presence of a minor planet at a distance of 0.003~au, which is within the typical Roche sphere radius. Hence, this minor planet is thought to be composed of dense, strong material, and is likely the remnants of a planetary core instead of a rocky asteroid.

\subsection{Theoretical considerations on the formation of debris}

The combined observations from transit photometry, infrared excess and emission line spectroscopy paint a rich and complex picture of planetary debris evolution inside of the Roche sphere. Consolidating this information into a unified theory for formation and evolution of the debris is challenging partly because of the variety of physics which must be considered.

{\rev Five} possible destruction outcomes exist for pollutants which are larger than dust \citep{broetal2017,mcdver2021}: (1) shearing apart through overspinning just outside of the Roche sphere \citep{makver2019,veretal2020a}, (2) sublimating partially or fully before reaching the Roche sphere \citep{steckloff2021}, (3) {\rev breaking up through mutual collisions of tidal fragments before reaching the Roche sphere \citep{broetal2022}}, (4) breaking up upon reaching the Roche sphere, or {\rev (5)} more rarely, barrelling through the traditional Roche sphere and impacting onto the photosphere of the white dwarf. Some white dwarfs might only encounter one pollutant over 10 Gyr of cooling, whereas other white dwarfs might receive, e.g., many pollutants in quick succession all on the same day, and then again for thousands of times over the next Gyr \citep{jura2008,wyaetal2014}.

The manner of approach and entry into the Roche sphere varies for different pollutants. Because the critical engulfment distance for giant stars is on an au-scale, and only dust and pebbles may be radiatively dragged to the Roche sphere, the traditional assumption was that any pollutants larger than pebbles needed to enter the Roche sphere on a highly eccentric orbit. However, this thinking was challenged with the discovery of the minor planet breaking up around WD~1145+017 in what appears to be a near-circular orbit \citep{rapetal2016,guretal2017,veretal2017a}. The first explanation posed to resolve this conundrum was by \cite{ocolai2020} who utilized tidal theory and ram pressure drag to demonstrate that circularization is possible before entering the Roche sphere.

Now a wider variety of entry methods into the Roche sphere is recognized. Dust and pebbles may enter on either a near-circular or highly eccentric orbit. The former is achieved through radiative drag alone \citep{veretal2022} or with the help of freshly generated gas \citep{malamud2021} or a strong magnetic field \citep{hogetal2021,zhaetal2021}. The latter can be prompted by gravitational interactions and lead to an extant disc capturing the pollutants \citep{griver2019}.

Asteroid, comet, moon and planet polluters similarly are not required to enter the Roche sphere on eccentric orbits because tides, ram pressure drag, and Ohmic drag can first circularize the orbits of these objects \citep{broken2019,veretal2019a,ocolai2020}. The possibility also exists of minor planets forming afresh from existing white dwarf discs \citep{vanetal2018} on near-circular orbits just outside of the Roche sphere; these would not be far away enough to escape eventual destruction back inside of the Roche sphere. Large polluters that do enter the Roche sphere on eccentric orbits do so at an angle which is dependent on the size of the perturber \citep{veretal2021}.

Despite the diversity of possibilities, the specific case of a solar system-like asteroid entering a white dwarf Roche sphere has received significant attention. The dynamics of the breakup and the evolution of the resulting fragments has been investigated with an increasingly sophisticated set of numerical and analytical tools \citep{jura2003,debetal2012,veretal2014b,veretal2015b,veretal2017a,duvetal2020,malper2020a,malper2020b,lietal2021,broetal2022,broetal2023a,broetal2023b}. These investigations have helped set the initial conditions for the subsequent evolution of the debris.

\subsection{Theoretical considerations on the evolution of debris}

After having formed into a disk-like or ring-like structure, the debris can persist for a wide variety of times (yrs to Myrs) before actually accreting onto the white dwarf \citep{giretal2012,verhen2020}. The debris orbiting WD~1145+017 on a 4.5-hour period has exhibited different photometric transit curves each year since its discovery. {\rev Long-lasting debris dics could also allow intact embedded planetesimals to migrate within the disc when certain conditions are met \citep{veretal2023b}}.

How the evolution proceeds is complicated by the dual presence of dust and gas. Their coupling admits limited analytical means to investigate their evolution. Hence, one strategy amongst researchers has been to decouple the dust and gas and focus on particular physical effects due to each. By focussing on dust, \cite{kenbro2017a} investigated collisional evolution, and \cite{faretal2017b} investigated magnetically-charged particles. By focussing on gas, \cite{mirraf2018} investigated how internal pressure can drive precession of disk-like structures, \cite{rozetal2021} demonstrated how gas can mechanically erode boulder-like debris, and \cite{treetal2021} illustrated how the persistent injection of gas from a disrupting planet or planetesimal can maintain an eccentric disc structure.

These investigations have usefully characterized different aspects of the physics of planetary debris evolution within a white dwarf's Roche sphere. However, the most accurate estimates of the rate of accretion onto the white dwarf photosphere itself require a coupled treatment of gas and dust because all dust will eventually sublimate before reaching the photosphere. Early coupled treatments \citep{bocraf2011,rafikov2011a,rafikov2011b,rafgar2012,metetal2012} emphasized the role of radiative drag on the eventual accretion rate, whereas later efforts have incorporated more collisional evolution \citep{kenbro2017b}, {\rev ongoing gas generation \citep{malamud2021}} and re-condensation \citep{okuetal2023} in their computations. 

\newpage

\acknowledgements

\subsection*{Acknowledgements}
{\rev We thank Uri Malamud, James Bryson and Judith Korth for their careful reading of the manuscript and for their expert advice, which have led to improvements}. A.J.M. acknowledges funding from the Swedish Research Council (grants 2017-04945 and 2022-04043) and the Swedish National Space Agency (grant 120/19C), and A.B. acknowledges support from a Royal Society Dorothy Hodgkin Research Fellowship, DH150130 and a Royal Society University Research Fellowship, URF{\textbackslash}R1{\textbackslash}211421.

\end{document}